\documentclass[twocolumn,twoside]{article}
\usepackage{graphics}
\newcommand{\hb}{\\ \hspace*{2ex}}

\begin{document}
\title{ABOUT THE GLOBAL MAGNETIC FIELDS OF STARS}
\author{V.D.\,Bychkov$^1$, L.V.\,Bychkova$^1$, J.\,Madej$^2$\\[2mm]
\begin{tabular}{l}
 $^1$ Special Astronomical Observatory of RAS,\hb
 Nizhnij Arkhyz, Russia, {\em vbych@sao.ru}\\
$^2$ Warsaw University Observatory, Warsaw, Poland {\em jm@astrouw.edu.pl}\\[2mm]
\end{tabular}
}
\date{}
\maketitle

ABSTRACT.
We present a review of observations of the stellar longitudinal (effective)
magnetic field ($B_e$) and its properties. This paper also discusses contemporary
views on the origin, evolution and structure of $B_e$.
\\[1mm]

{\bf 1. Introduction}\\[1mm]
At present there are collected direct measurements of the longitudinal (effective)
magnetic fields in 1873 stars of various spectral types. The total number of the
magnetic field $B_e$ measurements amounts to 24124. 
In the following text we shall refer to $B_e$ as the magnetic field for brevity.

\begin{figure}[h]
\resizebox{\hsize}{!}{\includegraphics{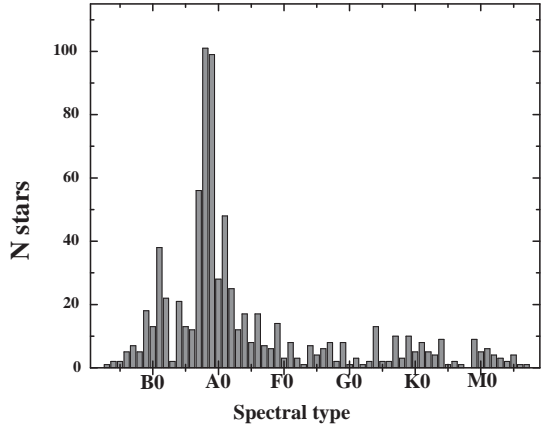}}
\caption{Number distribution of stars with measured longitudinal magnetic fields $B_e$ vs.
       spectral type.}
\end{figure}

The dominant part of existing observations (for over 900 objects) was obtained for CP stars.
\\[2mm]

{\bf 2. Observational data}\\[1mm]

We list here the most obvious advantages of the above progress: \\
1. There is accumulated a large set of $B_e$ measurements. \\
2. In some cases new magnetic measurements were obtained from spectra
       of relatively low resolution. \\
3. Those data were accumulated during a long time period (over 60 years),
       which actually allows one to study the long-period magnetic behavior
       of some objects. \\

\begin{table}[h]
\caption{Principal methods of $B_e$ measurements: }
\begin{tabular}{lr}
\hline
Method       &  N measurements \\
\hline
Phot.        &  5375 \\
Elc.         &  6991 \\
LSD and WDLS &  4083 \\
BS           &  1544 \\
FORS1/2      &  2540 \\
\hline
\end{tabular}
\end{table}

``Phot.'' stands for the photographic method (Babcock 1947a,b, 1958 and many others).
This method is now obsolete and is not used. \\
The ``Elc.'' method is an analogue of the photographic method,
but a CCD matrix is used as the receiver of light. 
Previously CCD matrix replaced a photographic plate in classical spectrometers. 
Currently echelle spectrometers are routinely used due to limited size of CCD matrices.
This method is still sometimes applied. \\ 
``LSD and WDLS'': It is a well known method, cf. Donati et al. (1997), 
Wade et al. (2000) and many other papers. This is a precise method,
which was actively in recent years and has yielded many new results.\\ 
``BS'' denotes the average surface field of stars. Such a number of measurements
does not imply that ``BS'' was measured for high number of stars. For some
slowly rotating CP stars BS was measured many times.\\
FORS1/2 stands for the low-resolution spectropolarimeter at the ESO
Very Large Telescope.\\
``H-line'' denotes $B_e$ measurements observed in hydrogen lines (Borra 
and Landstreet 1980, and many other papers).

\begin{figure}[h]
\resizebox{\hsize}{!}{\includegraphics{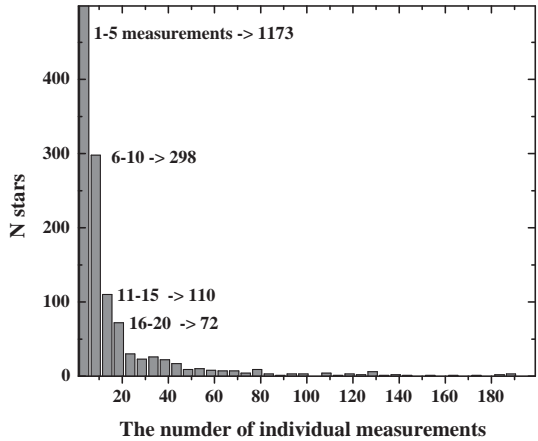}}
\caption{Number of individual $B_e$ measurements. }
\end{figure}

\begin{figure}[h]
\resizebox{\hsize}{!}{\includegraphics{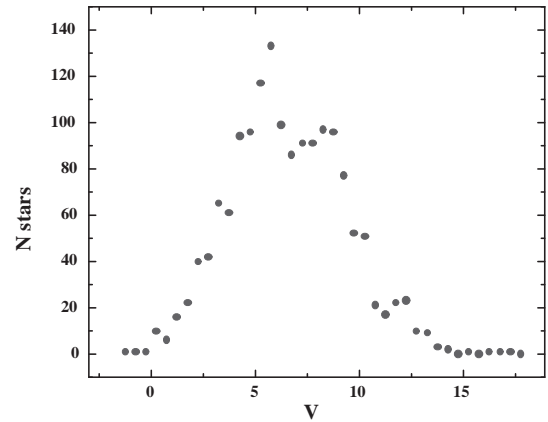}}
\caption{ Distribution of magnetic stars vs. apparent stellar magnitude.}
\end{figure}

\begin{figure}[h]
\resizebox{\hsize}{!}{\includegraphics{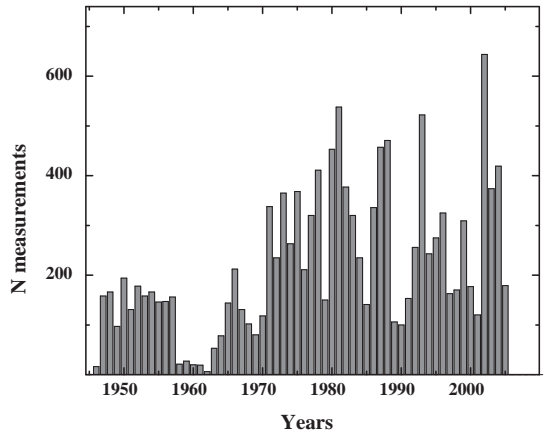}}
\caption{  Number of $B_e$ measurements obtained in various years.}
\end{figure} 

{\bf 3. Stars with known magnetic phase curves.}\\[1mm]

There exist 218 stars with measured phase curves of their longitudinal (effective)
magnetic field $B_e$. In that group, 172 objects are classified as magnetic chemically
peculiar stars. Remaining objects are stars of various spectral types, from the most
massive hot Of?p supergiants to low-mass red dwarfs and stars with planets.

\begin{table}[!t]
\centering
\caption{Number of stars for which magnetic phase curves were determined
   vs. the most important types. }
\label{list of stars} \tabcolsep1.2mm
\begin{tabular}{|l|r|}
\hline
All stars with mag. phase curves   &  218  \\
\hline
mCP stars                &  172  \\
\hline
Ae/Be Herbig stars       &    7  \\
\hline 
Be stars                 &    7  \\
\hline 
Supermassive Of?            &    3  \\
\hline 
Normal early B stars        &    5  \\
\hline
Flare stars                 &    3  \\ 
\hline 
TTS (T Tau type)            &    2  \\ 
\hline 
var. Beta Cep type          &    6  \\
\hline 
SPBS                        &    3  \\
\hline 
var. BY Dra type     &    4  \\ 
\hline 
var. RS CVn type     &    1  \\ 
\hline 
Semi-regular var.    &    1  \\ 
\hline 
DA                   &    1  \\
\hline 
var.pulsating stars  &    2  \\ 
\hline 
HPMS (high proper motions stars)      &    3  \\ 
\hline
var.Ori type         &    2  \\
\hline
\end{tabular}
\end{table}

Some stars were simultaneously put into two different classes.
For example, HD 96446 belongs to both the He-r and $\beta$ Cep
classes and HD 97048 belongs to both the TTS and Ae/Be
Herbig classes. The binary system DT Vir consists of two
companions: UV+RS (Flare + RS CVn type stars). Therefore,
the distribution of stars between classes had to be arbitrary or
redundant in some cases. \\
For example, Fig.5 shows the magnetic phase curve for mCp stars $\beta$
CrB. Periodic variability of the magnetic field of stars was described 
in more detail by Bychkov et al. (2005, 2013).

\begin{figure}[h]
\resizebox{\hsize}{!}{\includegraphics{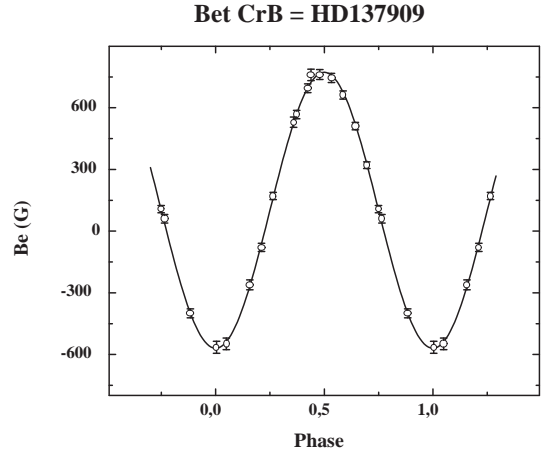}}
\caption{ Magnetic rotational phase curve of the mCp star $\beta$ CrB (HD 137909)
for the accurate rotational period derived by Wade et al. (2000).}
\end{figure}  

We selected the following most important conclusions about
the magnetic activity among stars of various types.
 
\begin{itemize}
\item{1.}
New class of magnetised objects was recently discovered -- supermassive
hot stars, type Ofp?. These stars show periodic variations of the longitudinal
magnetic field. Amplitudes of magnetic phase curves (MPC) reach several
hundred G. Of?p stars apparently are slow rotators. Configuration of their
magnetic field is represented by an oblique rotator.
\item{2.}
Magnetic fields were found among chemically normal early B stars. MPC's
were obtained for 3 stars of this type. In one object, HD 149438, MPC shows
complicated double wave shape, displayed also by some mCP stars. 
\item{3.} 
Magnetic field and its behaviour was best investigated in the group of 
mCP stars. Longitudinal magnetic fields $B_e$ have simple dipole 
configuration in majority of mCP stars (in 86 \% objects). Rotational 
magnetic phase curves often display simple harmonic shape with amplitudes
reaching 10 kG.

Remaining 14 \% of investigated mCP stars display more complex phase 
curves being a superposition of two sine waves and have either dipole 
or more complex structure of their global magnetic fields. Amplitudes 
of rotational $B_e$ variation essentially do not differ from those in 
``sine-wave'' mCP stars.

\item{4.} 
Solar-type stars have global magnetic fields of low strength, seldom
approaching few dozens of G. Measuring of such low-intensity fields meets
with many methodologicacl difficulties. Therefore, we can only suppose,
that in some investigated stars (in $\xi$ Boo A, for example) magnetic
phase curves appear as simple harmonic waves. Very significant 
progress in measuring of magnetic fields in stars was achieved using 
the ZDI method (magnetic cartography of the surface). More credible
considerations require higher number of investigated stars and still
higher accuracy of magnetic field observations. Moreover, it is known 
that magnetic properties of solar-type stars vary periodically in time 
scale from few years to several dozens of years.

\item{5.}
Ae/Be Herbig stars usually exhibit magnetic rotational phase curves 
of a purely harmonic shape with amplitudes reaching several hundred G.

\item{6.} 
Magnetic phase curves of pulsating $\beta$ Cep stars 
vary with the period of rotation. MPC show a complicated structure
with low amplitudes of dozens G. Closely related slowly pulsating B stars
(SPB) also display longitudinal magnetic field varying with the period of
rotation. MPC show a simple harmonic shape with amplitudes reaching several
dozens G.

\item{7.} 
T Tau stars have magnetic fields of complex structure, display also 
complex magnetic phase curves with amplitudes approaching several
hundred G. Undoubtedly, fields of such a strength have to strongly
influence accretion of matter onto stars.

\item{8.}
Late-type stars -- M dwarfs have global magnetic fields of complex
structure. Magnetic rotational phase curves only roughly can be 
approximated by a superposition of two waves. This was also directly
confirmed by recent observations using the ZDI method. Amplitudes of 
variations of the integrated longitudinal magnetic fields reach
several hundred G. Some stars present an amazing feature, stepwise
creation or anihilation of the global magnetic field and related
$B_e$ variations.
 
\item{9.}
HD 189733 -- this is a typical dwarf of spectral class K2V, where a 
giant planet, ``hot Jupiter'' was found. Central star in the system 
is a solar-like object. The star possesses magnetic field which is 
typical for its spectral class, and its longitudinal component 
varies with the amplitude of several G.

\end{itemize}

{\bf 4. mCp stars.}\\[1mm]

Magnetic fields of stars are best studied for mCp stars. 
One of major problems for these stars is the relations between their magnetic field
and the chemical composition.
We proposed a way to clarify this problem (Bychkov et al. 2009).
We defined relative magnetization (MA) for different types of chemically peculiarity
comparing distributions of their occurrence with the observed $<B_e>$.
Example of such a distribution for stars of Si peculiarity is shown in Fig. 6.
Number distribution of CP stars vs. $T_{\rm eff}$ for all
different types of chemical peculiarity was shown in Fig. 7.
Magnetization ``MA'' for various subclasses of CP stars vs.
$T_{\rm eff}$ was shown in Fig. 8.
Reduction of ``MA'' with the reduction of $T_{\rm eff}$ is apparent there
for H-r, He-w and Si stars.
Such a reduction of ``MA'' supports the fossil theory of the magnetic
field origin in those stars. 
If the age of a star is high, then its mass is lower and ``MA'' also is lower.
But we see sharp rise of ``MA'' about $T_{\rm eff} = $ 10000 $K^o$.
Therefore, we raise the assumption that the dynamo mechanism joins at this
point on the $T_{\rm eff}$ scale.

\begin{figure}[h]
\resizebox{\hsize}{!}{\includegraphics{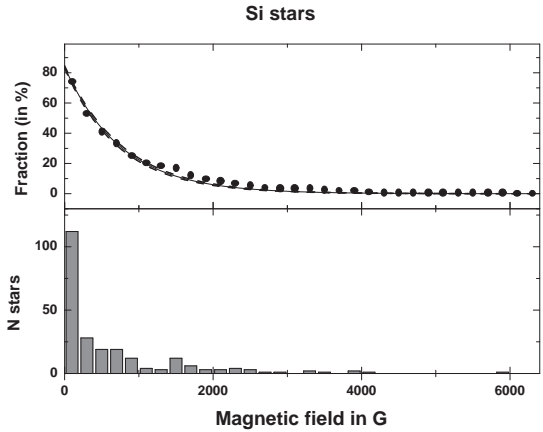}}
\caption{Integrated distribution function $N_{Int}(B)$ in percent
         (upper panel), and the number distribution function $N(B)$ (lower
         panel) for stars of Si peculiarity type. }
\end{figure}  

\begin{figure}[h]
\resizebox{\hsize}{!}{\includegraphics{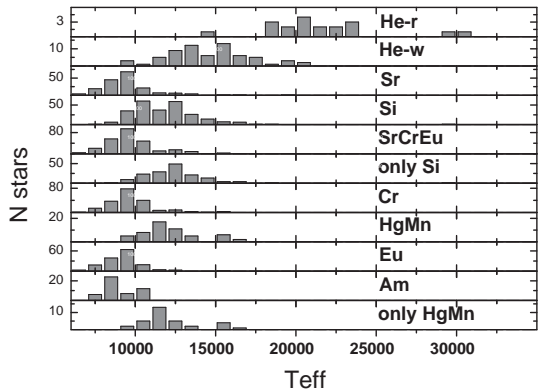}}
\caption{ Number distribution of CP stars vs. $T_{\rm eff}$ for 
         various types of chemical peculiarity.}
\end{figure}  

\begin{figure}[h]
\resizebox{\hsize}{!}{\includegraphics{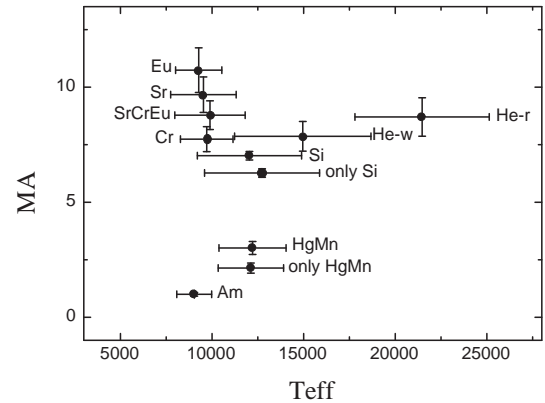}}
\caption{Magnetization (MA) for various subclasses of CP stars.
          Bars define the range of $T_{\rm eff}$ and MA occupied by a
          given subclass. }
\end{figure}  

{\bf Summary.}\\[1mm]

In recent years significant progress was attained in the study of stellar
magnetism. While previously one could measure and discuss behaviour of the 
stellar magnetic field only in mCP stars, white dwarfs and the Sun, currently 
we can measure and collect data on the magnetic field for many more types of  
stars ranging from supermassive hot giants to fully convective cold dwarfs of
low mass. One can note significant contribution of the MiMeS collaboration
which has discovered a new class of magnetic objects, supermassive hot giants
Ofp? type and other magnetised hot stars. These discoveries significantly
extended our knowledge about magnetism of hot stars and in future will give
rise to our understanding of processes in stellar atmospheres and 
circumstellar space.

One can expect that rapid accumulation of new observational data will allow
one to study in detail the variability of stellar magnetic field in stars
both of different spectral types and evolutionary stages. We share the
conviction that the magnetic field and its evolution is a crucial 
agent of stellar physics.

\unitlength=1in
{\it Acknowledgements.} We acknowledge support
from the Polish Ministry of Science and Higher Education grant
No. N N203 511638 and the Russian grant ``Leading Scientific
Schools'' N4308-2012.2.
\\[3mm]

\indent
{\bf References\\[2mm]}
Babcock H.W.: 1958, {\it Ap.J.Suppl.Ser.,} {\bf 30}, 141. \\
Borra E.F., Landstreet J.D.: 1980, {\it Ap.J.Suppl.Ser.,}
       {\bf 42}, 421. \\
Donati J.F., Semel M., Carter B.D., Rees D.E., Cameron A.C.: 1997,
               {\it MNRAS,} {\bf 291}, 658. \\
Wade G. A., Donati J.-F., Landstreet J. D., Shorlin S. L. S.: 2000, 
        {\it MNRAS,} {\bf 313}, 823. \\
Babcock H. W.: 1947a, {\it ApJ,} {\bf 105}, 105. \\
Babcock H. W.: 1947b, {\it PASP,} {\bf 59}, 260. \\
Bychkov V.D., Bychkova L.V. and Madej J.: 2005, {\it A\&A,} {\bf 430}, 1143. \\
Bychkov V.D., Bychkova L.V. and Madej J.: 2013, {\it AJ,}  {\bf 146:74}, 10pp. \\

\vfill
%
\end{document}